\documentclass{emulateapj}

\newcommand{\unit}[1]{\ifmmode\,{\rm #1}\else$\,{\rm #1}$\fi}

\newcommand{\phnm}{$\phantom{-}$}
\newcommand{\eg}{{\it e.g.}\ }
\newcommand{\ie}{{\it i.e.}\ }
\newcommand{\etal}{~{et~al.}\ }  
\newcommand{\ROSAT}{{\it ROSAT}\ }
\newcommand{\xmm}{{\it XMM$-$Newton}}
\newcommand{\Chandra}{{\it Chandra}\ }
\newcommand{\msun}{M_{\odot}}
\newcommand{\lsun}{L_{\odot}}
\newfont{\gwpfont}{cmssq8 scaled 1000}
\newcommand{\rexcess}{REXCESS}
\newcommand{\rfive}{$r_{500}$}
\newcommand{\w}{$\langle w \rangle$}

\usepackage{natbib}
\slugcomment{Accepted 2010 Feb 17}

\shorttitle{BCGs and Core Gas}
\shortauthors{Haarsma\etal}

\begin{document}

\title{Brightest Cluster Galaxies and Core Gas Density in REXCESS Clusters}
\author{
Deborah B.\ Haarsma\altaffilmark{1},
Luke Leisman\altaffilmark{1},
Megan Donahue\altaffilmark{2},
Seth Bruch\altaffilmark{2},
Hans B\"{o}hringer\altaffilmark{3}, 
Judith H. Croston\altaffilmark{4},
Gabriel W. Pratt\altaffilmark{5,3},
G. Mark Voit\altaffilmark{2},
Monique Arnaud\altaffilmark{5},
Daniele Pierini\altaffilmark{3}
}

\altaffiltext{1}{Calvin College, 1734 Knollcrest SE, Grand Rapids, MI
49546, dhaarsma@calvin.edu}

\altaffiltext{2}{Physics and Astronomy, Michigan State University,
East Lansing, MI 48824-2320}

\altaffiltext{3}{Max-Planck-Institut f\"ur extraterrestriche Physik,
Giessenbachstra{\ss}e, 85748 Garching, Germany}

\altaffiltext{4}{School of Physics and Astronomy, University of
Southampton, Southampton, SO17 1BJ, UK}

\altaffiltext{5}{Laboratoire AIM, DAPNIA/Service d'Astrophysique -
  CEA/DSM - CNRS - Universit\'{e} Paris Diderot, B\^{a}t. 709,
  CEA-Saclay, F-91191 Gif-sur- Yvette Cedex, France}

\begin{abstract}

We investigate the relationship between brightest cluster galaxies
(BCGs) and their host clusters using a sample of nearby galaxy
clusters from the Representative \xmm\ Cluster Structure Survey
(\rexcess).  The sample was imaged with the Southern Observatory for
Astrophysical Research (SOAR) in R~band to investigate the mass of the
old stellar population.  Using a metric radius of 12\unit{h^{-1}}~kpc,
we found that the BCG luminosity depends weakly on overall cluster mass
as $L_{BCG} \propto M_{cl}^{0.18\pm0.07}$, consistent with previous
work.  We found that 90\% of the BCGs are located within 0.035~\rfive\
of the peak of the X-ray emission, including all of the cool core (CC)
clusters.  We also found an unexpected correlation between the BCG
metric luminosity and the core gas density for non-cool core (non-CC)
clusters, following a power law of $n_{e} \propto L_{BCG}^{2.7\pm0.4}$
(where $n_e$ is measured at 0.008~\rfive).  The correlation is not
easily explained by star formation (which is weak in non-CC clusters)
or overall cluster mass (which is not correlated with core gas
density).  The trend persists even when the BCG is not located near
the peak of the X-ray emission, so proximity is not necessary.  We
suggest that, for non-CC clusters, this correlation implies that the
same process that sets the central entropy of the cluster gas also
determines the central stellar density of the BCG, and that this
underlying physical process is likely to be mergers.

\end{abstract}


\section{Introduction}
\label{intro}

The intracluster gas in some galaxy clusters shows a central
concentration in a cool core.  The higher density of this core gas
allows more rapid energy loss in the form of X-ray emission.  In the
absence of other energy sources, the classic ``cooling flow'' model
suggests that the central gas should cool.  The cool gas would rapidly
condense and form stars at rates greater than
100$\msun$\unit{yr^{-1}}, but optical colors and spectral lines
indicate much lower rates of less than $\sim$10\unit{\msun yr^{-1}}
(reviewed in \citealp{donahue04a}).  Thus, the gas must be heated by
other processes, such as AGN activity or other forms of feedback that
regulate the thermal properties of the gas (reviewed in
\citealp{mcnamara07a}).  Yet the identity of the feedback process is still
widely debated, and current simulations tend to overpredict the
fraction of clusters with cool cores (\eg \citealp{kay07a}).

The gas core is often located near the Brightest Cluster Galaxy (BCG),
particularly in cool core clusters
\citep{jones84a,bildfell08a,rafferty08a,sanderson09b}.  The BCG is not
only luminous and massive, but has a more extended optical light
profile than other ellipticals, due to its unique merger history at
the bottom of the gravitational potential well (\eg\
\citealp{hausman78a,vale08a}).  Models predict that the BCG mass is
correlated with the total cluster mass (\eg \citealp{delucia07b}), but
the observed correlation is typically weaker
\citep{lin04a,popesso07a,brough08a,whiley08a,yang08a,mittal09a}.

To better understand the connection between BCGs and cool cores, we
investigated the BCGs in 31 southern hemisphere clusters from the
Representative \xmm\ Cluster Structure Survey (\rexcess)
\citep{boehringer07a}.  The \rexcess\ sample was chosen to evenly
represent the range of X-ray luminosities in the local ($0.06<z<0.18$)
population, with no bias regarding X-ray morphology or central surface
brightness.  Thus, it provides an ideal laboratory for testing the
connections between BCGs and their host clusters.  \cite{pratt07a},
\cite{croston08a}, and \cite{pratt09a} measure the X-ray
luminosity, temperature, mass, core gas density, and cooling time.
Here we present ground-based CCD (optical R~band) imaging of the
cluster BCGs and calculate colors relative to 2MASS K~band magnitudes.
We investigate the connections between the old stellar population and
the properties of the X-ray gas.

Interestingly, the strongest correlation we found was not between the
BCG mass and the total cluster mass, but between the central BCG
stellar density and the core gas density.  We describe the cluster
sample and X-ray measurements in \S\ref{x.obs}, and the R~band
observations and BCG magnitude calculations in \S\ref{opt}.  In
\S\ref{results} and \ref{special} we examine this new correlation, as
well as other connections between BCGs and their host clusters.  Our
conclusions are summarized in \S\ref{summary}.  We use $h=0.7$,
$\Omega_m=0.3$, and $\Omega_\Lambda=0.7$ throughout.

\section{X-ray Sample}
\label{x.obs}

The \rexcess\ sample \citep{boehringer07a} contains 33 clusters
selected to evenly sample a range of X-ray luminosities and
temperatures (2 to 9~keV), including a variety of dynamical states and
core cooling times.

\cite{croston08a} derived the radial gas density profiles for the
\rexcess\ sample from \xmm\ surface brightness profiles, using a
non-parametric deprojection and PSF-deconvolution
method. \cite{croston06a} showed that this method accurately recovers
the gas density profile observed in higher resolution \Chandra data, down
to $2\farcs4$ radius, for data of similar statistical quality to
\rexcess.  Two clusters, RXCJ0956 and RXCJ2152, were excluded from the
X-ray analysis because they have multiple distinct components which
preclude a one dimensional profile analysis, leaving a sample of 31 clusters.

Relevant X-ray properties are listed in Table~\ref{tab.xray}.  The
scaling radius \rfive\ (defined as the radius enclosing a mean
overdensity of 500 times the critical density) and the corresponding
cluster mass $M_{cl}$ was found from iterations about the
$M_{500}-Y_{X}$ relation of \cite{arnaud07a}.  The cooling time was
determined at a radius of 0.03~\rfive\ (about 10$''$-20$''$).  The
core gas density $n_e$ was measured at a radius of 0.008~\rfive\
(about 3$''$-6$''$) and was scaled by $h(z)^{-2}$ to remove evolution
effects.  The dynamical state of the gas was characterized by \w, the
standard deviation of the offsets $w$ between the X-ray peak and the
centroids of emission found for various radii (B{\"o}hringer\etal, in
prep).



\cite{pratt09a} identified a subsample of \rexcess\ as cool core
clusters using a cutoff in core gas density.  We chose instead to use
a cutoff in cooling time (at 0.03~\rfive), since the cooling time can
be compared to various dynamical time scales in the cluster.  We
defined cool core (CC) clusters to be those with $t_{cool} < 2$~Gyr
(log $t_{cool} < 9.3$), marked with blue stars on the figures.
Non-cool core (non-CC) clusters have $t_{cool} > 2$~Gyr.  Compared to
\cite{pratt09a}, this definition shifts only one cluster into the CC
sample, namely RXCJ0211, see \S\ref{special}.

\cite{pratt09a} also identified clusters as morphologically disturbed
if \w$>0.01$\rfive.  We use the same cut-off but with revised values
of \w from B{\"o}hringer\etal (in prep).  Disturbed clusters are marked
with red squares on the figures.  In these clusters the gas
distribution is less symmetric, likely due to a recent merger of
sub-clusters and a younger dynamical age.  In general, disturbed
clusters do not have cool cores (although RXCJ1302 and RXCJ2319 are
exceptions).

For each cluster, the position of the X-ray peak emission has been
determined by B{\"o}hringer\etal (in prep), who describe the method
in detail.  Briefly, the peak position for most clusters is the local
maximum in a 4$''$ smoothed image.  In a few clusters the central
surface brightness is flat, so the peak position is given as the
center of a dipole fitted at 0.1~\rfive.  The positions were confirmed
by visual inspection.  These positions are more precise and accurate
than the low-resolution \ROSAT positions listed in
\cite{boehringer07a}.

\section{Optical data}
\label{opt}

\subsection{Observations}
\label{opt.obs}

Table~\ref{tab.obs} lists our observations, which, with the exception
of two targets, were made with the 4.1~m Southern Observatory for
Astrophysical Research (SOAR) in Cerro Pach{\'o}n, Chile.  Most of the
SOAR observations used the SOAR Optical Imager (SOI,
\citealp{walker03a,schwarz04a}), consisting of two thinned,
back-illuminated 4096x4096 CCDs.  The field of view is over 5$'$
across, with a binned pixel size of 0$\farcs$154 from a natural pixel
size of 0$\farcs$0767.  The standard observing strategy was three
200~s exposures, dithered by 10$''$ to span the gap between the two
CCDs.  The remaining SOAR observations used the Goodman Spectrograph
\citep{clemens04a} in imaging mode, which utilizes a single Fairchild
CCD. The field of view is 7$\farcm$2 in diameter, with a binned pixel
size of 0$\farcs$29 from a natural pixel size of 0$\farcs$15.  The
standard observing strategy was a single 600~s exposure.  One target
(RXCJ1044) was observed at the 42~inch (1.1~m) John S. Hall Telescope
at Lowell Observatory in New Mexico, using a 2048x2048 CCD camera with
pixel size 0$\farcs$59; we made 15 dithered exposures of 600~s each.
All observations were made under
clear photometric conditions; if the conditions or data were suspect
in any way, the galaxy was reobserved.  The seeing was typically 1$''$.
Finally, one target (RXCJ1311) was mistakenly omitted from
observations but is one of the few to appear in the Sloan
Digital Sky Survey (SDSS, \citealp{adelman-mccarthy08a}).  SDSS data
have sufficient resolution (pixel size 0$\farcs$396) and depth for our
measurements; a Luminous Red Galaxy spectral template was used to
transform the $r'$ data to the Cousins R filter (center $\sim$630~nm)
that we used for all other observations.  

The images were reduced in the standard fashion using IRAF v.2.12.2,
including bias subtraction, flat field correction, and cosmic ray
removal.  The dithered images were aligned and median combined for
each target, compensating for differing background levels if
necessary.  The background value of the combined image was measured
using APPHOT in a 1 to 2 arcminute radius annulus centered on the BCG
with the radius chosen to avoid field objects.   
The data were corrected for airmass assuming an R~band extinction
coefficient of $-0.1$, and calibrated to the Vega magnitude scale
using \cite{landolt92a} standard stars or \cite{hamuy92a}
spectrophotometric standard stars.  The data were converted to the AB
scale with $R_{AB} = R_{vega} + 0.206$; all magnitudes and colors are
reported on the AB scale.

The uncertainty in the final magnitudes is dominated by systematics.
Because the BCGs are bright and cover many pixels, the random error on
the summed flux and on the subtracted background is much smaller than
the total systematic error in the flat-field corrections, standard
star calibration, airmass corrections, and variation in sky conditions
throughout the night.  We estimate the uncertainty due to these
systematics to be around 0.05~mag and assume this value for all
observations.

Images of a select subsample of the BCGs are shown in
Figure~\ref{fig.mosaic1}.  The entire sample is available at {\tt
http://www.calvin.edu/$\sim$dhaarsma/rexcess}.  The optical images are
overlayed with contours of the \xmm\ data, taken in the
$0.5-2$~keV~band with all three detectors combined.  The X-ray surface
brightness images were corrected for vignetting and 
detector gaps, point sources were removed and smoothly
refilled, and the images were smoothed with a 4$''$ Gaussian.


\subsection{Surface brightness profiles}
\label{opt.profile}

Figure~\ref{fig.profiles} shows the observed surface brightness
profiles of the BCGs, corrected for Galactic extinction, to restframe,
and for cosmological surface brightness dimming. The slopes appear
similar across the sample.  

\cite{postman95a} characterized the profile using the slope of the
integrated profile, $\alpha = d\log(L_r)/d\log(r)$.  They found that
$\alpha$ depends on metric luminosity, and that this dependence could
be removed to improve the utility of BCGs as standard candles.  We
measured $\alpha$ for this sample and found no correlation with metric
luminosity, as well as an absence of low $\alpha$ values.  This
confirms the work of P.~Lyman discussed in \cite{boehringer01a}. 

We also fit de~Vaucouleurs functions to the profiles, but again found
no correlation between profile shape ($r_{\rm eff}$) and metric
luminosity.  (A typical de~Vaucouleurs function is plotted in
Figure~\ref{fig.profiles}.)  Instead, we found the metric luminosity
to be correlated with surface brightness at a given metric radius (the
vertical offset in Figure~\ref{fig.profiles}).  Thus, these BCGs share
the same profile shape but differ in their overall brightness.

\subsection{BCG identification}
\label{opt.id}

\begin{figure}
\plotone{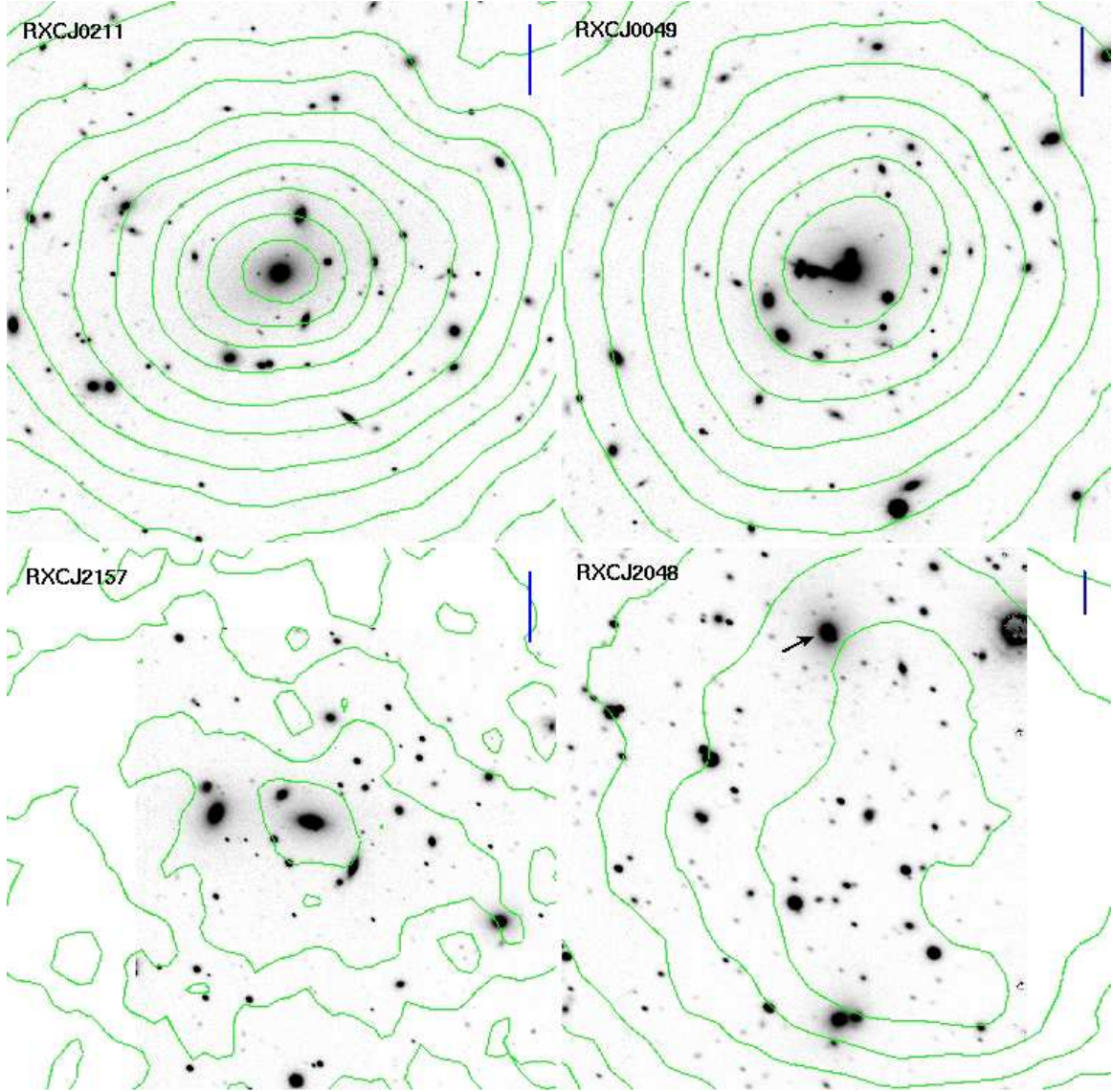}
\caption{Images of \rexcess\ BCGs.  The complete sample can be viewed
at {\tt http://www.calvin.edu/$\sim$dhaarsma/rexcess/}.  The four
clusters shown here illustrate our classification scheme: RXCJ0211 is
type 1 (brightest, smooth), RXCJ0049 is type 2 (brightest, lumpy),
RXCJ2157 is type 3 (multiple galaxies of similar brightness) with the
X-ray peak clearly selecting one galaxy, and RXCJ2048 is type 3 with
the X-ray peak far from any galaxy.  North is up and east left.
Greyscale: R~band. Vertical bar: 50~kpc at redshift of cluster.
Contours: X-ray surface brightness from \xmm, $0.5-2$~keV, with
contours increasing in steps of $\sqrt{2}$.
\label{fig.mosaic1}
}
\end{figure}

\begin{figure}
\epsscale{0.9}
\plotone{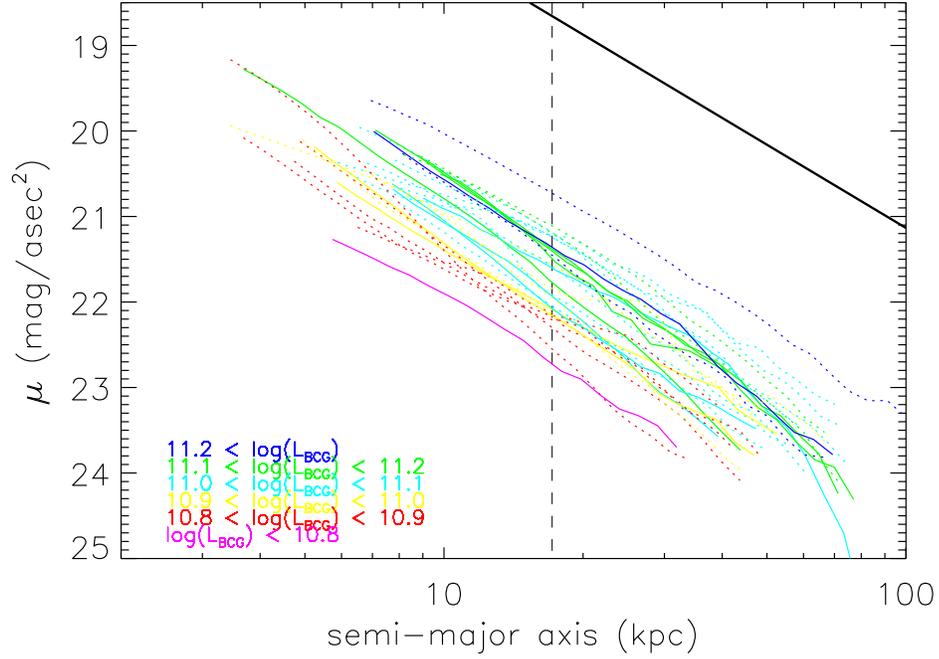}
\caption{R band surface brightness profiles of the BCGs, including
k-corrections, extinction corrections, and corrections for surface
brightness dimming.  The data are plotted only for radii greater than
3~arcsec and less than twice the isophotal radius (isophote of 22.5
mag\unit{arcsec^{-2}}). Solid lines are the cool core population, 
dotted lines are the remaining BCGs.  The vertical dashed line
indicates the metric radius of 12$h^{-1}$~kpc.  The thick line in the
upper right indicates the shape of a typical de~Vaucouleurs profile 
(effective radius of 60~kpc).  In the online version of the paper,
the colors of the lines indicate
the metric luminosity of the BCG, showing the correlation with the
surface brightness at the metric radius.  
\label{fig.profiles}
}
\end{figure}

Tables~\ref{tab.opt-met} and \ref{tab.opt-iso} list the optical
properties of the BCGs.  We classified the clusters into three
categories.  1) Clusters where a central galaxy is clearly the
brightest and has a generally smooth light distribution.  2) Clusters
where a central galaxy is clearly the brightest, but has multiple
light peaks appearing in the BCG envelope. These may be stars,
foreground galaxies, infalling galaxies on a radial path, member
galaxies in low radius orbits, or mergers in progress.  3) Clusters in
which there are two or more galaxies of similar brightness.
Figure~\ref{fig.mosaic1} includes examples of each type.

We identified the BCG using aperture magnitudes, with two borderline
cases decided by proximity to other galaxies or to the X-ray peak.
The fields were inspected visually in the Digital Sky Survey out to
500\unit{h^{-1}}~kpc; if there was any ambiguity about the brightest
galaxy or the X-ray peak position, the 5$'$-7$'$ field of view was
positioned to cover other candidates.  The brightness was measured for
galaxies within this field of view (typically a 200\unit{h^{-1}}~kpc
radius around the X-ray peak). We used an unmasked 50\unit{h^{-1}}~kpc
aperture in order to include light from the extended envelope and
mergers in progress.  One borderline case was RXCJ2234, in which three
galaxies had nearly the same brightness (within 0.01 mag) and none
were near the X-ray peak, so we choose the one located in the middle
of the group.  The other case is RXCJ1311 (Abell1689), in which some
galaxies had a neighbor galaxy within the 50\unit{h^{-1}}~kpc aperture
which boosted their brightness; measurements in a smaller
(12\unit{h^{-1}}~kpc) unmasked aperture identified two galaxies with
similar brightness (within 0.06 mag) near the center of the group, so
we choose the one located at the X-ray peak.

BCGs located more than 0.03~\rfive\ from the X-ray peak are marked
with an arrow in Figure~\ref{fig.mosaic1} and the online images, and
with a green X in Figures~\ref{fig.color-mag}-\ref{fig.nevclustm}.
Table~\ref{tab.opt-iso} lists the 2MASS identification of the galaxy,
but we were able to better resolve the crowded fields and determine a
more accurate position of the BCG center, which we list in
Table~\ref{tab.opt-met}.

\subsection{Integrated Magnitudes}
\label{opt.mag}

Because of the crowded fields around BCGs, a simple aperture magnitude
would be biased by extra flux from other galaxies in the
field. Instead, we masked light from sources unrelated to the BCG,
then fit elliptical isophotes to the remaining light distribution
using the IRAF task ELLIPSE.  We then integrated the isophotes
numerically to get the total flux.

Masked objects included stars, foreground galaxies, and cluster
members not physically near the BCG.  Some of the brightness peaks in
the envelope, however, are the cores of cluster members in the process
of merging with the BCG (such multiple nuclei are particularly visible
in BCGs we classify as type 2). One approach would be to include the
light from all objects that are already merging in the envelope.  That
is difficult, however, to implement, because it would require
discriminating between foreground galaxies and galaxies physically
near the BCG, and determining {\it a priori} how much a galaxy must be
apparently assimilated in order to include it with the BCG light.  We
chose instead to mask all brightness peaks except the BCG center, and
to fit isophotes only to light in a smooth elliptical distribution
around that center.  This procedure provides a consistent measure
without subjective decisions on which brightness peaks to include.  As
a test, we measured the magnitudes in a 50\unit{h^{-1}}~kpc radius
aperture without any masking, and found that the standard deviation of
the absolute magnitudes increased from 0.29 to 0.34 mag; thus, the
small masked aperture is a more consistent measure.

When fitting ellipses, we fixed the central coordinates but allowed
the ellipticity and position angle to vary with radius.  If the
ellipticity or position angle changed rapidly or the fit failed due to
too much masking, we adjusted the masks to achieve a smooth fit.  To
prevent double counting we discarded any overlapping isophotes.

The fitted isophotes were then integrated numerically, effectively
replacing flux in the masked areas.  During integration, we simulated
a circular metric aperture by calculating the fraction of each ellipse
inside the circular aperture.  The apparent metric magnitude in a
radius of 12$h^{-1}$~kpc (17~kpc for our assumed cosmology, about
5$''$-15$''$) is listed in Table~\ref{tab.opt-met} as $R_{met}$.

The apparent magnitude was converted to an absolute magnitude using
the known redshifts.  We applied the Galactic extinction corrections
of \cite{schlegel98a} as implemented in the calculator at the NASA
Extragalactic Database; corrections ranged from 0.04 to 0.40 mag.  We
converted the magnitudes to rest-frame using the {\it kcorrect}
software \citep{blanton07a}, assuming a non-evolving Luminous Red
Galaxy spectral template; corrections range from 0.05 to 0.18 mag.  In
Tables~\ref{tab.opt-met} and \ref{tab.opt-iso} we list magnitudes and
colors with and without k-corrections so workers may apply their own
SED and star formation history.

\subsection{Colors}
\label{opt.color}

BCGs typically have similar colors because they have old stellar
populations.  To confirm this, we measured $R-K$ colors for the
sample, making use of the Two Micron All Sky Survey Extended Source
Catalog (2MASSX, \citealp{2mass}).  In 2MASSX, the recommended catalog
magnitudes ($K_{k20fe}$) are isophotal ($K=20$~mag\unit{arcsec^{-2}}
with typical radii 10 to 50~kpc) inside an elliptical aperture with
bright neighbors masked out.  To determine accurate colors, however,
both bands must be measured in the same aperture with the same objects
masked.  We could either use the R~band mask and ellipses
(\S\ref{opt.mag}) on the 2MASS image (giving a $K$ to match
$R_{met}$), or use the 2MASSX mask and ellipses on the R~band image
(giving an $R_{K20}$ to match $K_{k20fe}$).  We used the latter method
because of the difference in pixel scale between SOAR and 2MASS
(0$\farcs$154 vs. 1$''$).  If the precise SOAR masks were applied to
the low resolution 2MASSX image, many more pixels of BCG light would
be masked, leaving few pixels for the ellipse fit.  Instead, we masked
the same objects in the SOAR image as were masked in 2MASSX and fixed
the central coordinates, ellipticity, and position angle to the 2MASSX
values during the ellipse fits.  We then proceeded with integration 
as in \S\ref{opt.mag}.  The resulting $R_{K20}$ values and
$R-K$ colors are listed in Table~\ref{tab.opt-iso}.

Figures~\ref{fig.color-mag} and \ref{fig.color-color} show the $R-K$
colors as they depend on absolute magnitude and on the $J-K$ color
from 2MASSX.  With the exception of RXCJ1302 and RXCJ2319 (see
\S\ref{special}), the colors have very little scatter and no trend
with abolute magnitude.  This consistency demonstrates that our
methods have not introduced significant systematic errors relative to
the 2MASS photometry.

\section{Correlation of BCG and Gas properties}
\label{results}

\subsection{Cluster mass vs. BCG mass}
\label{mass-mass}

Hierarchical formation models suggest that the mass of the BCG should
be correlated with the mass of its host cluster.  For example,
\cite{whiley08a} recently calculated the expected dependence, using
the models of \cite{delucia07b}, to be $M_{BCG}\propto M_{cl}^{0.4}$
or $M_{cl}^{0.5}$ depending on the feedback model used.  A correlation
between BCG luminosity and cluster X-ray luminosity has long been
noted in the literature (\eg\ \citealp{schombert88a, edge91a, edge91b,
hudson97a}).  Here we give an overview of recent results, which tend
toward a shallower power law dependence than predicted by models.
For example, \cite{whiley08a} measured the correlation as
$M_{BCG}\propto M_{cl}^{0.12\pm0.03}$ for K~band magnitudes inside a
diameter of 37~kpc (radius of 13$h^{-1}$~kpc).  \cite{brough08a} saw a
similar dependence of $L_{BCG}\propto M_{cl}^{0.11\pm0.10}$ at K~band
inside 12$h^{-1}$~kpc.  \cite{yang08a} found $L_{BCG}\propto
M_{cl}^{0.17}$ for galaxy groups.  \cite{lin04a} found a steeper
dependence ($L_{BCG}\propto M_{cl}^{0.26\pm0.04}$), but they used
isophotal magnitudes rather than metric magnitudes. (Isophotal
magnitudes favor larger, more massive galaxies; in terms of
Figure~\ref{fig.profiles}, isophotal magnitudes make a horizontal cut
through the profiles while metric magnitudes make a vertical cut.)
Similarly, \cite{popesso07a} found $L_{BCG}\propto M_{cl}^{0.25}$
using r~band Petrosion magnitudes.  \cite{stott08a} used K~band
isophotal magnitudes and found an even steeper relation ($-1.1\pm0.3$
mag per decade of X-ray luminosity, or $L_{BCG}\propto
L_{cl}^{0.44\pm0.12}$) when fitting only to clusters with high X-ray
luminosity.  \cite{mittal09a} used K~band total magnitudes
(extrapolating the Sersic profile) and found $L_{BCG}\propto
M_{cl}^{0.62\pm0.05}$; they note that total magnitudes give a stronger
correlation with cluster properties than smaller aperture magnitudes.

For our sample, we measured $L_{BCG}$ in R~band inside 12$h^{-1}$~kpc.
We confirm the weak trend between BCG luminosity and cluster
mass. Figure~\ref{fig.mass-mass} shows the trend, which has
substantial scatter; Table~\ref{tab.stats} shows that the correlation
is marginally significant.  Unlike \cite{stott08a}, we do not see the
trend becoming more prominent at higher X-ray luminosities.  Fitting
in log-log space using the bivariate correlated errors and intrinsic
scatter (BCES(Y$\vert$X)) regression method \citep{akritas96a}, we
find $L_{BCG}\propto M_{cl}^{0.18\pm0.07}$, consistent with the
results found by others using metric magnitudes.

\subsection{Core gas density vs. BCG ellipticity}
\label{ellipticity}

We noticed a possible correlation between core gas density and BCG
ellipticity.  Neglecting RXCJ2319 (see \S\ref{special}), the largest
ellipticity among the cool core clusters is 0.29.  The largest
ellipticity among the non-CC clusters is 0.53, and half of the non-CC
clusters have ellipticities larger than 0.29.  This comparison
suggests that BCGs in CC clusters are rounder than those in
non-CC clusters.  A K-S test shows that this trend is not
statistically significant for the small numbers of our sample, so a
larger sample is needed to test the trend suggested by these data.

\subsection{Core gas density vs. BCG location }
\label{xrayoff}

The offset between the X-ray peak and the BCG position is listed in
Table~\ref{tab.xray} and plotted in Figure~\ref{fig.nevxrayoff}.  The
\xmm\ resolution prevents a precise measure of this offset at the
smallest separations, but \xmm\ can clearly identify clusters where the
BCG is located far from the X-ray peak.  We find that three BCGs are
located more than 0.035~\rfive\ (and more than 30$''$) from the X-ray
peak, all of which have multiple galaxies of similar brightness.  That
leaves 28/31 (90\%) of the clusters with a galaxy closer than
0.035~\rfive.  Note that the \rexcess\ sample is unbiased with
respect to X-ray morphology, so this close agreement is not due to an
overrepresentation of bright X-ray cores in the sample.
 
Our high fraction of small offsets (90\%) is higher than \cite{lin04a}
found in a sample of 93 clusters: $\sim$65\% within the same radius
($0.02~R_{200}$ in their Figure 1).  The difference from our results is
likely because their sample is infrared-selected, has X-ray data drawn
from several catalogs, and uses X-ray peak positions from lower
resolution \ROSAT data.

We also found more small offsets than \cite{loubser09a} found in a
sample of 49 BCGs.  They found 45\% of BCGs falling within 20~kpc of
the X-ray peak while we find 77\% within that distance.  This
difference is likely due to different sample selection and to their
exclusion of galaxies that fall near the peak if the peak is far from
the literature X-ray position (their Table 7).

Our sample is more comparable to that of \cite{sanderson09b}, who
study the LoCUSS sample of 65 X-ray selected clusters.  They found
75\% of BCGs falling within 0.04~\rfive\ of the X-ray centroid, while we
found 90\% within that distance of the X-ray peak.  These percentages
are consistent given Poisson statistics.  

Our results agree with \cite{hudson09a}, who study 64 clusters in the
HIFLUGCS sample. They found 88\% to have a BCG within 50$h^{-1}$~kpc
of the X-ray peak, while we find 90\% within that radius.

Recent work by \cite{bildfell08a}, \cite{rafferty08a}, and
\cite{sanderson09b} found that star formation in cool cores requires
that the X-ray and galaxy centroids lie within $\sim$20~kpc of each
other.  Our data agree; all 11 CC clusters have offsets less than
20~kpc, while 13 out of 20 non-CC have such small offsets.  Like
\cite{sanderson09b}, we find that steep central X-ray profiles have
small BCG offsets, similar to the correlation between central gas
density with BCG offset shown in Figure~\ref{fig.nevxrayoff}.

\cite{bildfell08a} find a correlation between BCG absolute B~band
magnitude and the offset from the X-ray peak, where BCGs are fainter
when located far (up to 500~kpc) from X-ray peak.  We see no such 
correlation at large offsets in R~band.

\subsection{Gas dynamical state vs. BCG location }
\label{wvxrayoff}

We noticed a possible correlation between large values of \w\
(indicating disturbed gas) and large offset
between the BCG and the X-ray peak.  RXCJ2048 and RXCJ2129 are two of
the three most disturbed clusters (the other is RXCJ2157), and are
also two of the three largest BCG offsets (the other is RXCJ2234).
The correlation is difficult to quantify for small offsets, since many
of our measured offsets are consistent with zero.  The correlation for
large offsets is not surprising: clusters with disturbed gas have
likely undergone recent mergers and the largest galaxies have not had
time to settle into the core.

\subsection{Core gas density vs. BCG central stellar density in non-CC clusters} 

\subsubsection{Observed Trend}
\label{nevstarm}

We compared several optical properties of BCGs with various X-ray
properties of the intracluster gas, and in most cases we saw little or
no correlation.  (Some of these are reported in Table~\ref{tab.stats}
and describe above.)  The tightest correlation, however, between X-ray
and optical properties was one we did not expect: between the core gas
density of non-CC clusters and their aperture BCG luminosity, shown in
Figure~\ref{fig.nevstarm}.

The correlation appears to follow a power law, with outliers mainly
toward higher core gas density. The rank order correlation statistics
are reported in Table~\ref{tab.stats}: for the non-CC sample
(excluding RXCJ0211 and RXCJ2048, see \S\ref{special}), the parameters
show less than 0.16\% chance of being uncorrelated, a detection
stronger than 3 sigma.  A BCES(orthogonal) fit in log-log space finds
$n_e \propto L_{BCG}^{2.7\pm0.4}$.  We also note that the correlation
cannot be a spurious effect of the X-ray point spread function, since
non-CC clusters have broad, flat centers that are well resolved by
\xmm.

Note that the correlation is with aperture luminosity, not total
luminosity.  The aperture luminosity is proportional to the mass in
the inner 12$h^{-1}$~kpc, and so is a good proxy for the central
stellar mass density of the galaxy (it is not a good proxy for the
total mass of the galaxy).  When discussing this trend, we will refer
to $L_{BCG}$ as central stellar density.  Thus, we discovered that,
for non-CC clusters, the BCG central stellar density is larger in
clusters with larger core gas densities.  We tested the correlation
using several measures of BCG luminosity, but found that well-masked
metric magnitudes give the tightest correlation (Table~\ref{tab.stats}
shows P=0.0016).  Well-masked isophotal magnitudes in R~band also
showed some correlation (P=0.015), but $K_k20fe$ isophotal mags from
2MASSX showed little correlation (P=0.14). 2MASSX reports magnitudes
with some bright objects masked, but not as completely masked as we
could do with deeper, better-sampled images in R~band.

We tested whether other related parameters would show the relationship
more clearly, but found that core gas density and BCG metric
luminosity had the strongest correlation (Table~\ref{tab.stats}).  In
place of BCG metric luminosity, we tried metric surface brightness.
In place of core gas density, we tried cooling time and centroid
shift.  In no case was the correlation as tight as the one
shown. \cite{croston08a} reported the correlation of $n_e$ with X-ray
centroid shift \w\ (in opposing directions for the CC and non-CC
samples), which in turn introduces a correlation of $L_{BCG}$ with \w,
but this is not as tight as $L_{BCG}$ with $n_e$.  We tested if the
correlation of $n_e$ and $L_{BCG}$ could be tightened further by
removing any additional dependence on \w, but saw little improvement.

We note that the correlation is driven by a population of 5 clusters
with the lowest core gas densities: RXCJ0145, 0225, 2129, 2157, 2234.
These 5 have not only the lowest core gas density, but are also the
only clusters with multiple galaxies of similar brightness (which we
classify as type 3). Three of the clusters (RXCJ0145, 2129, 2157) are
among the most disturbed clusters in the sample.  All three indicators
suggest a young dynamic age, so this subsample points to cluster
dynamics as key to the interpretation of the relation.

\subsubsection{Interpretation}
\label{nevstarmtheory}

What could be the physical cause of this correlation?  In no way do we
suggest that the ICM gas density has a {\it direct} effect on the
stars of the BCG; that would be unphysical.  Rather, the explanation
of the correlation must be based in some other physical process that
affects both the core gas density and the central stellar density.
What processes could be involved?

BCGs and core gas are known to be related via the stars that form from
the cooling gas.  This causes BCGs in CC clusters to show more
indicators of star formation than BCGs in non-CC clusters, including
H$\alpha$ emission (\eg\ \citealp{cavagnolo08b}), blue color (\eg\
\citealp{bildfell08a}), and younger stellar populations (\eg\
\citealp{loubser09a}).  Similar studies are being done for the REXCESS
sample (Donahue\etal submitted) Our correlation, however, is seen in
R~band, which detects the mass of the old stellar population, not
recent star formation.

In fact, it seems unlikely that this old stellar population should be
correlated with the presence or absence of a cool core, since the cool
core gas and the stars have such different time scales.  Prominent
cool cores have radiative cooling times shorter than a billion
years, whereas the vast majority of the stars formed several
billion years ago, based on studies of their color-magnitude relation
and spectral energy distributions at $z\sim1$ (\eg
\citealp{andreon08a, stott08a}).  \cite{collins09a} suggest that even
the assembly of BCGs is mostly complete before redshift 1. 

For these reasons, the correlation is not explained by the
relationship between strong cool cores and star formation.  Moreover,
the correlation is in the non-cool core population, not the cool core
population.  In non-CC clusters, the gas is much too diffuse to cool
and form stars.  Some process other than star formation must be at
work.

The correlation is not due to overall cluster properties, such as
total mass, X-ray luminosity, or gas temperature.  It is not the case
that more massive clusters simply have more luminous BCGs and denser
core gas.  While the BCG central stellar density does depend weakly on
total mass (Figure~\ref{fig.mass-mass}) with a power of $0.18\pm0.07$,
that trend cannot solely explain the much stronger dependence (power
of $2.7\pm0.4$) on core gas density.  Moreover, the core gas density
has no correlation with system temperature \citep{croston08a} or with
total cluster mass (Figure~\ref{fig.nevclustm}, Table~\ref{tab.stats}).

The correlation is not primarily due to the BCG gravitational
potential.  It is true that the BCG is located near the X-ray peak in
most cases (Figure~\ref{fig.nevxrayoff}), and that a larger stellar
density would increase the gravitational potential well.  If this were
the sole cause, however, the core gas density would be expected to be
only modestly dependent on the central stellar density, but the
observed relationship has a power of 2.7.  In addition, two clusters
with BCGs located far from the X-ray core also fall on the trend,
implying that physical proximity is not required.

The correlation might be due to a few clusters that have experienced a
recent merger.  As noted above, the 5 clusters with lowest gas density
include the three clusters with the most disturbed gas, suggesting a
recent merger (Table~\ref{tab.stats} shows the strong correlation of
gas density and \w in nonCC clusters).  These 5 clusters also have
multiple galaxies of similar brightness rather than a single dominant
BCG, as expected for a recent cluster merger where the central
galaxies have not yet merged into one BCG.  After the merger, the
resulting BCG would have higher total luminosity, but before the
merger there is a time window when the gas density is low and the BCG
total luminosity is low.  This sounds like a promising explanation of
the trend, but there is a problem.  The $L_{BCG}$ we report is a
metric aperture luminosity in 12$h^{-1}$~kpc, not a total luminosity;
as noted above, it is a better proxy for central stellar density than
for total mass.  How is central stellar density affected by a merger?
Mergers between ellipticals are known to decrease the central density
cusp, producing the well-known trend of lower central surface
brightness in higher luminosity elliptical galaxies.  Thus, the
galaxies corresponding to low gas density are those with low stellar
densities, which would be {\it after} their mergers rather than the
cuspy galaxies seen in the window between gas merger and galaxy
merger.

We suggest that mergers are key to the explanation, but that it is the
long-term merger history of the cluster rather than the most recent
merger.  Mergers of sub-clusters cause gas shocks which increase the
gas entropy.  More massive clusters have larger shocks and thus larger
entropy gains.  In the absence of galaxies and feedback, models show
that the final entropy profile of the gas is determined by total
cluster mass, leading to a self-similar family of entropy profiles
(reviewed in \citealp{voit05a}).  The entropy at the core, however,
tends to deviate from the self-similar relation.

The core gas density can serve as a proxy for the deviation of the
core entropy relative to other clusters of the same mass.  In a
self-similar cluster model, the ratio of core entropy $K_0$ to the
entropy $K_\Delta$ at some large scale radius $r_\Delta$ is
\[
      \frac{K_0}{K_{\Delta}} = \left( \frac{T_0    }{n_0^{2/3}}     \right) 
                         \left( \frac{n_{\Delta}^{2/3}}{T_{\Delta}} \right) .
\]
The gas density $n_{\Delta}$ at $r_\Delta$ is the same for all
clusters by definition.  The ratio $T_0/T_\Delta$ can be affected by
non-gravitational processes but is closely tied to gravitational 
potential, so $T_0/T_\Delta$ is not observed to vary much across
self-similar clusters.  Thus, in practice, the expression reduces to
\[
           \frac{K_0}{K_{\Delta}} \propto n_0^{-2/3}.
\]
In other words, the central gas density is closely related to the
relative enhancement of central entropy over what would be expected in
a self-similar cluster, independent of cluster mass.  A cluster whose
central entropy was unusually large for its halo mass would therefore
have an unusually low central density.

Many physical processes can change the core entropy level, including
radiative cooling, supernova or AGN feedback, and conduction.  What we
have discovered here is that, whatever the process, it also appears to
set the central stellar density of the BCG.  While most processes that
affect gas entropy have little effect on stars, mergers affect
both. The details of the merger history, such as the size of the
subclusters, may be the cause of these deviations from the
self-similar value.  As stated above, mergers also affect stellar
density, decreasing the cusp and reducing the central stellar density.
Thus, mergers could increase gas entropy, decrease gas density, and
decrease stellar density.  In this way, the central stellar density
would provide a record of the merger history of the cluster. The core
gas in non-CC clusters has a very long cooling time and does not
experience feedback, so it too preserves a memory of the shocks
experienced in the merger history.  In our sample, even disturbed
clusters (red squares on the figures) fall on the trend, suggesting
that the correlation is dominated by long-term merger history, not
recent merger events.

The CC sample does not follow a similar power-law trend. In these
clusters, radiative cooling and feedback have a greater influence on
the core gas, allowing the entropy to decrease and the gas density to
increase beyond the level set by mergers. This increase does not have
a particular correlation with BCG stellar density.  The outliers on
the upper side of the non-CC trend may be in transition to higher gas
density via these processes.

\section{Special Cases}
\label{special}

RXCJ0211: This cluster is classified by \cite{pratt09a} as non-CC
based on its central gas density and central cooling time, but we
classify it as CC based on its cooling time at 0.03~\rfive\ (see
\S\ref{x.obs}).  It has the lowest gas temperature of all the clusters
in \rexcess, giving it a shorter cooling time than other clusters of
similar gas density such as RXCJ0645.  Like all CC clusters in
\rexcess, it hosts a central radio source, something that some non-CC
clusters lack (a full analysis of \rexcess\ radio properties will
appear in Heidenreich et al, in prep).  If RXCJ0211 is instead put in
the non-CC sample, the trend between core gas density and BCG stellar
density (Figure~\ref{fig.nevstarm}, \S\ref{nevstarm}) is still present
but a bit less significant (R=0.57, P=0.01).

RXCJ0049 and RXCJ0225: In these clusters, the galaxy located at the
X-ray peak is in the 2MASS Point Source catalog but not in the 2MASSX
extended source catalog.  Thus, we did not have the K~band profile
parameters needed to measure the $R-K$ color using the method in
\S\ref{opt.color}.

\begin{figure}
\epsscale{0.9}
\plotone{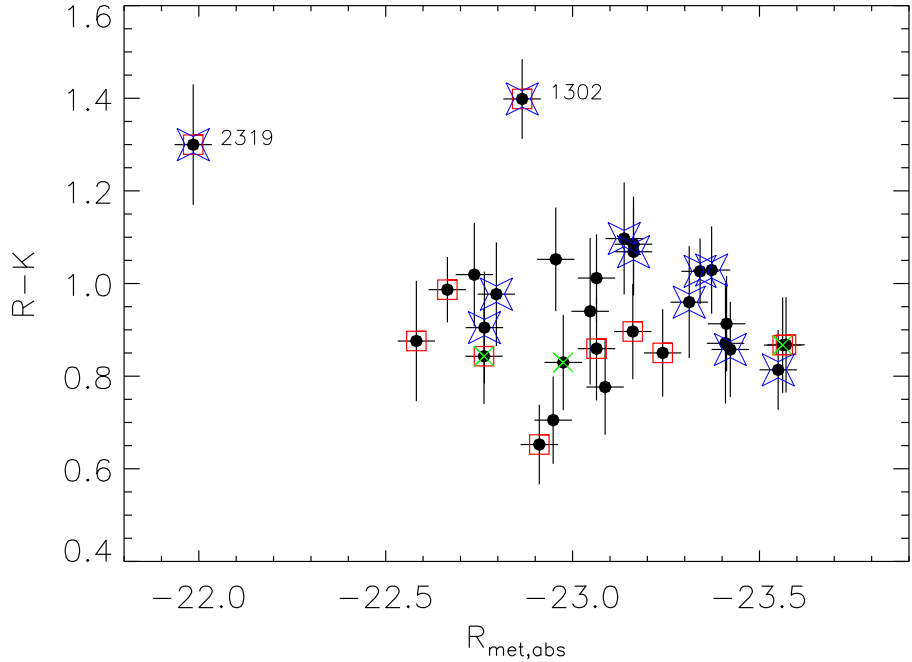}
\caption{BCG color vs. absolute metric magnitude in AB units,
restframe, corrected for Galactic dust extinction. Symbols indicate
special sub-samples: cooling time less than 2~Gyr (blue star),
disturbed X-ray emission (red square), and large separation between
BCG and X-ray peak (green X).  In general, the BCGs have similar color
and no dependence of color on magnitude; the exceptions are RXCJ1302
and RXCJ2319, see \S\ref{special}.
\label{fig.color-mag}
}
\end{figure}

\begin{figure}
\epsscale{0.9}
\plotone{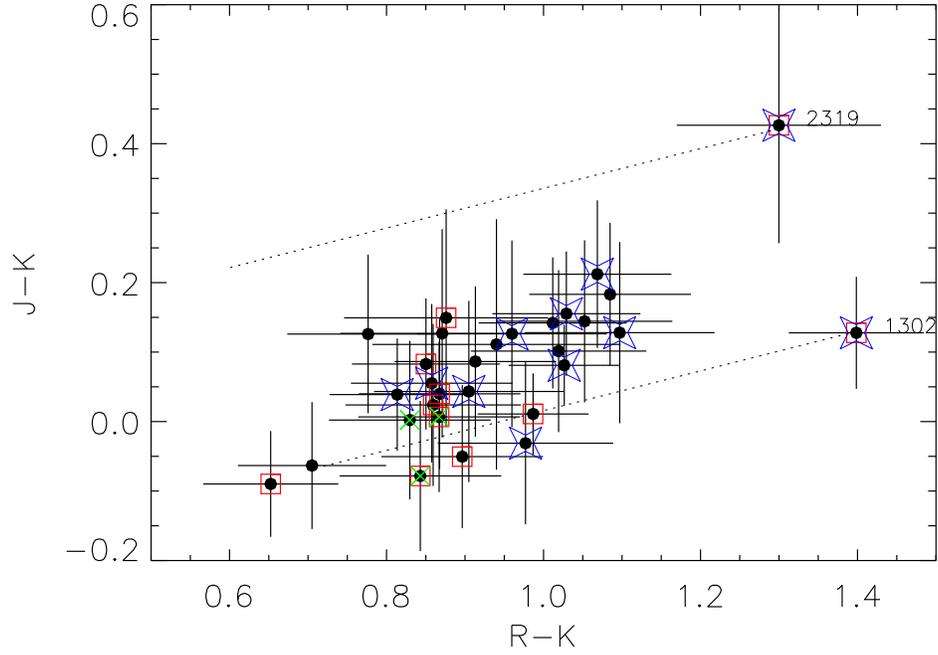}
\caption{BCG colors in AB units, restframe, corrected for Galactic
dust extinction. Symbols same as Figure~\ref{fig.color-mag}.  The
dotted lines indicate reddening due to dust internal to the BCG; the
length of the line indicates extinction from $A_V=0$ to 1 for the
outliers RXCJ1302 and RXCJ2319, see \S\ref{special}.
\label{fig.color-color}
}
\end{figure}

\begin{figure}
\epsscale{0.9}
\plotone{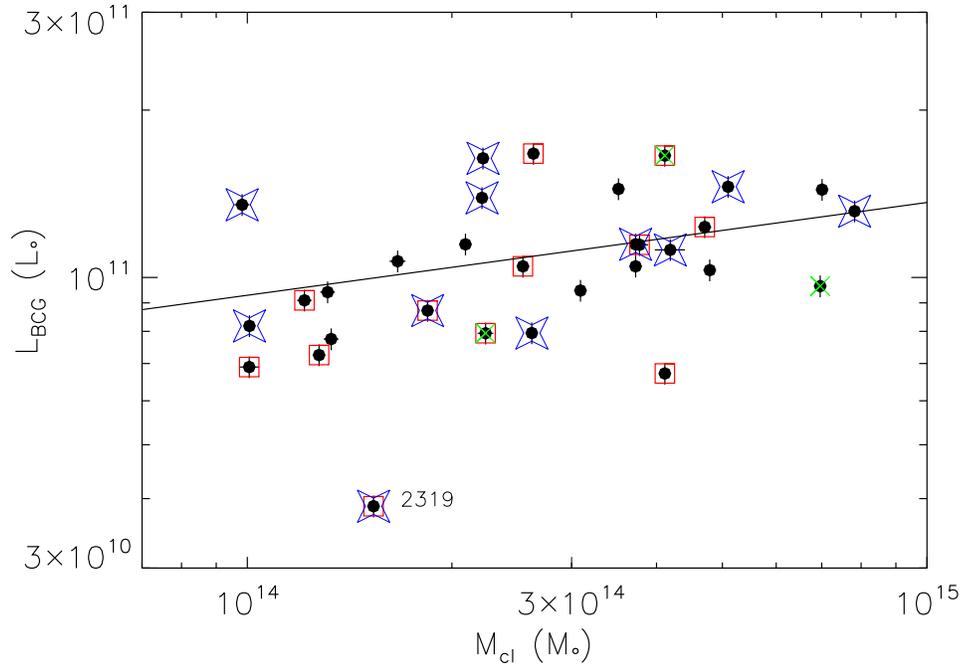}
\caption{The BCG metric luminosity (radius 12$h^{-1}$~kpc) depends
slightly on total cluster mass.  Symbols same as Figure~\ref{fig.color-mag}.
The line shows the BCES(Y$\vert$X) fit to the data (excluding RXCJ2319, see
\S\ref{special}) with a power law of $0.18\pm0.07$.
\label{fig.mass-mass}
}
\end{figure}

\begin{figure}
\epsscale{0.9}
\plotone{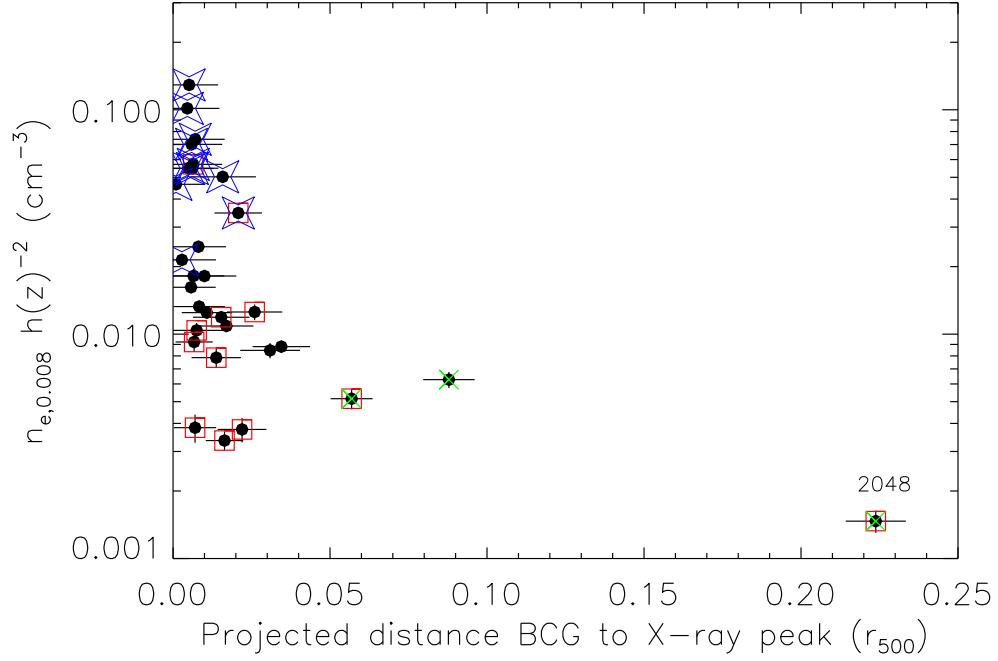}
\caption{In cool core clusters, the BCG is co-located with the peak of
the X-ray emission; 90\% of BCGs in the sample are located within
0.035~\rfive.  Symbols same as in Figure~\ref{fig.color-mag}.
\label{fig.nevxrayoff}
}
\end{figure}

\begin{figure}
\epsscale{0.9}
\plotone{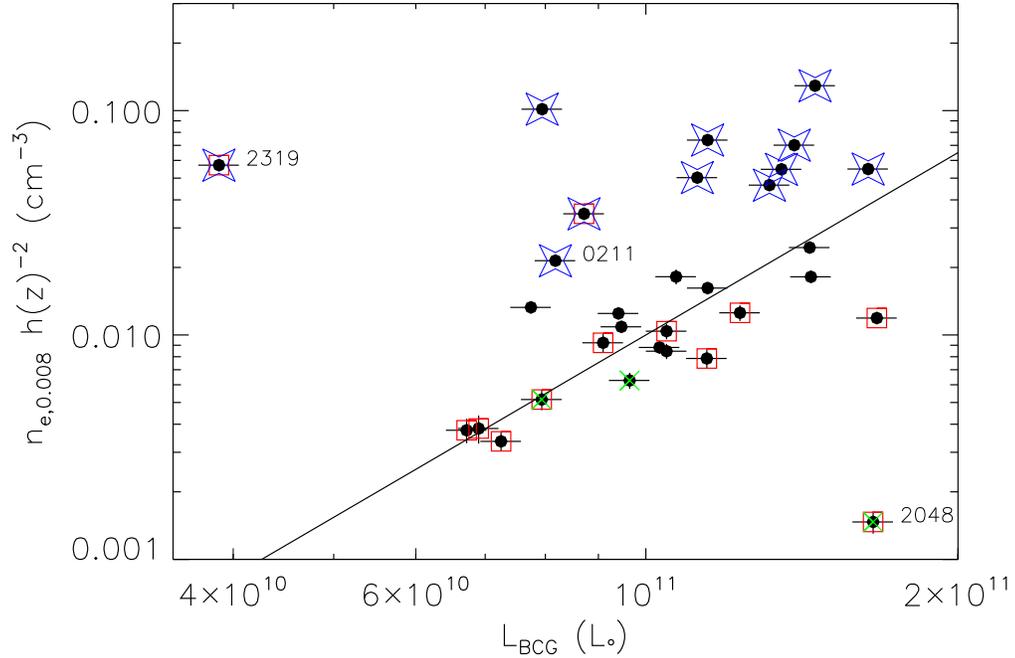}
\caption{The gas density of the cluster at 0.008~\rfive\ 
vs. the BCG metric luminosity (effectively the stellar mass density inside a
radius of 12$h^{-1}$~kpc).  Symbols same as Figure~\ref{fig.color-mag}.
The line is a BCES(orthogonal) fit to the non-cool core population
(excluding RXCJ2048, see \S\ref{special}), with a power law of
$2.7\pm0.4$.
\label{fig.nevstarm}
}
\end{figure}

\begin{figure}
\epsscale{0.9}
\plotone{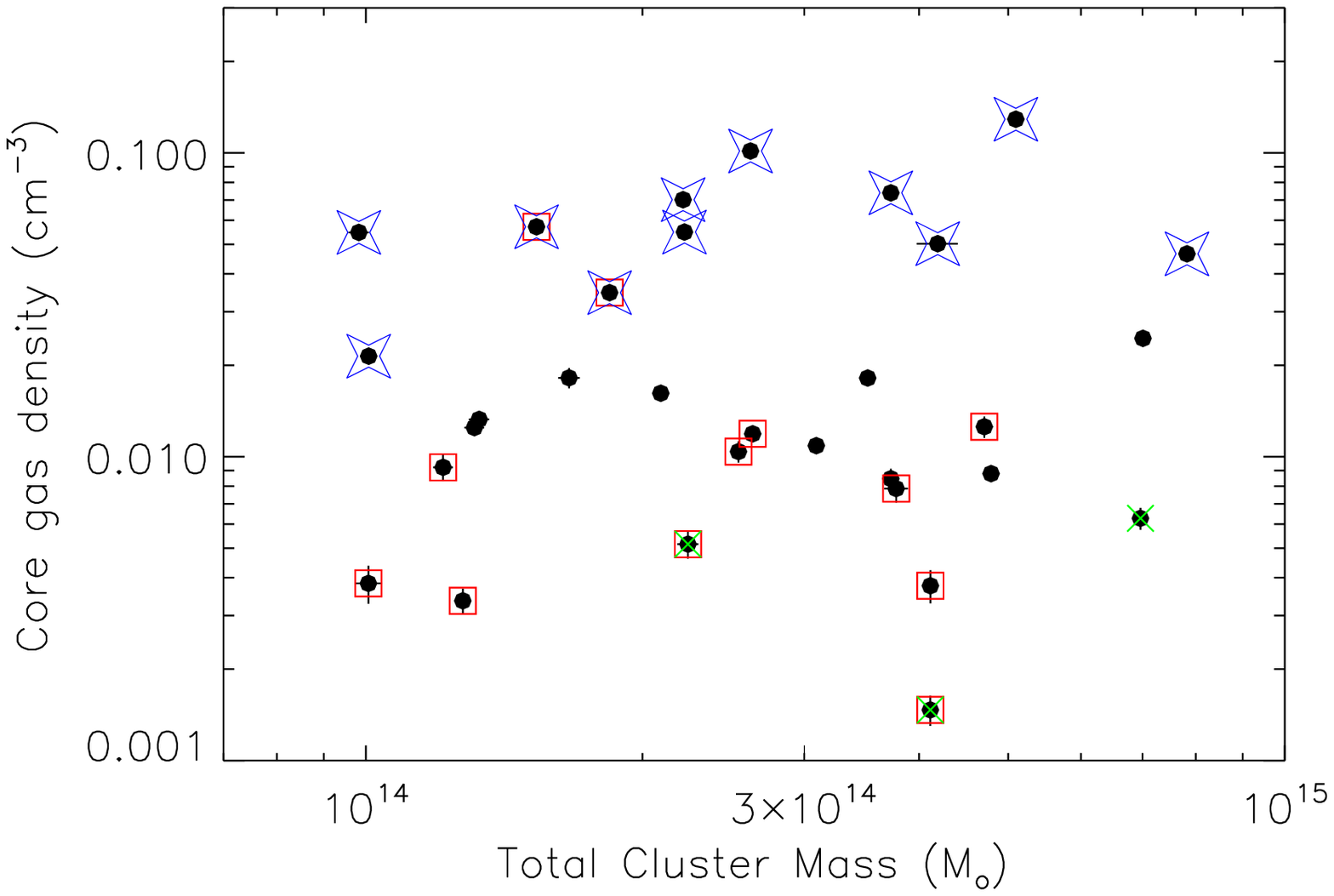}
\caption{The gas density of the cluster at 0.008~\rfive\ 
vs. the total cluster mass. Symbols same as Figure~\ref{fig.color-mag}.
There is no correlation between the core gas density and the
overall cluster mass. 
\label{fig.nevclustm}
}
\end{figure}

RXCJ1302: This galaxy has a redder $R-K$ color than the rest of the
BCGs in the sample (Figure~\ref{fig.color-mag}).  Yet its
$R_{met,abs}$ magnitude (Figure~\ref{fig.color-mag}) and $J-K$ color
(Figure~\ref{fig.color-color}) are consistent with the rest of the
sample.  One possible explanation is dust internal to the galaxy
(found in other BCGs in CC clusters, e.g. \citealp{egami06a,odea08a}).
Assuming a typical extinction curve \citep{gordon03a}, about 0.5 mags
of internal extinction in $R_{abs,met}$ would bring the $R-K$ color in
line with the rest of the sample while keeping $J-K$ consistent with
the sample.  If such a change is made, the power fit in
Figure~\ref{fig.mass-mass} would change only slightly to
$0.17\pm0.07$.  The fit in Figure~\ref{fig.nevstarm} is unaffected
because this cluster is not in the non-CC sample.

RXCJ2048: This cluster is an outlier in the lower left of
Figure~\ref{fig.nevstarm}.  Of the clusters in the sample, it has the
largest separation between the BCG position and X-ray peak
(Figure~\ref{fig.nevxrayoff}). The BCG may be misidentified, although
the only other candidate in the field is slightly fainter and located
just as far from the X-ray peak on the opposite side of the X-ray
centroid.  Its X-ray properties are unique as well, having the most
diffuse X-ray emission (Figure~\ref{fig.mosaic1}) and the longest
cooling time (Table~\ref{tab.xray}) of the entire sample.  It is
possible the gas is not even in hydrostatic equilibrium.  Because of
these multiple issues, we leave it out of the fit in Figure
\ref{fig.nevstarm} and out of the statistics in Table~\ref{tab.stats}.

RXCJ2319: This galaxy has a much fainter absolute magnitude than the
rest of the BCGs in the sample (Table~\ref{tab.opt-met}).  We
reobserved the galaxy on another night and confirmed the faint $R$
magnitude.  A misidentification is unlikely, since the galaxy is
clearly the brightest in the field and coincident with the X-ray peak.
A faint magnitude and red color could be caused by an incorrect
redshift, but the optical redshift $z=0.0984$ from \cite{guzzo09a} is
confirmed by our own observations of faint emission lines in the BCG
at $z=0.0979$ (Donahue \etal submitted).  In addition, its $R-K$ and
$J-K$ colors are extremely red (Figures~\ref{fig.color-mag},
\ref{fig.color-color}), while its $J-H$ color from 2MASSX is
consistent with the sample.  That suggests an excess of $K$ emission,
perhaps similar to the obscured Seyfert BCG in Abell 1068
\citep{edge02a}, except the optical spectrum shows very typical line
widths rather than the strong lines of AGN activity. Thus, we do not
have a good explanation for this object, but because of its puzzling
nature we leave it out of the fits in Figures~\ref{fig.mass-mass} and
\ref{fig.nevstarm} and out of the statistics in Table~\ref{tab.stats}.

\section{Summary}
\label{summary}

Using the \rexcess\ sample of galaxy clusters, we investigated the
relationships between the BCG stellar mass and the properties of the
X-ray emitting gas.  We confirmed the weak correlation seen by others
between the BCG luminosity and the total cluster mass, and the close
proximity of the BCG to the X-ray peak. 

We detected a trend among the non-CC clusters in which the core gas
density increases with the BCG aperture luminosity, a proxy for the
BCG central stellar mass density.  This trend is much clearer when the
BCG luminosity is determined from well-masked aperture magnitudes
rather than from unmasked or isophotal magnitudes.  This trend holds
even in cases where the gas is disturbed or the BCG is located far
from the central region.  We argue that the core gas density is an
indicator of the deviation of the central entropy from the
self-similar value.  Thus, the core gas entropy and the central BCG
stellar density appear to be more closely related than previously
thought.  We suggest that cluster mergers could be the underlying
cause, since mergers can both increase gas entropy and decrease BCG
central stellar density.  If so, this trend could set important
constraints on models of the central gas in clusters and of BCG
formation.

These results are based on a sample of only 31 clusters, and need to
be confirmed using other, larger cluster samples.  For example,
\cite{cavagnolo09a} recently fit entropy profiles to 239 clusters from
the \Chandra X-ray Observatory archive.  The optical data
corresponding to this or other large samples of X-ray selected
clusters would allow an investigation of the trend with a
statistically significant sample size. If the trend is real and our
interpretation is correct, it should be present independent of how the
clusters are selected.


\vskip .2in

\small
DBH acknowledges the support of a Calvin Research Fellowship. LRL
acknowledges the support of a Sid Jansma Summer Research Fellowship.
MD and GMV acknowledge the support of an LTSA grant NASA NNG-05GD82G.
MD and SB acknowledge the support of MSU research funds.  
DBH and LRL thank the AstrW10 class of Calvin College for
their assistance with observations of RXCJ1044. 

We present data obtained with the Southern Observatory for
Astrophysical Research, which is a joint project of Conselho Nacional
de Pesquisas Cientificas e Technol\'ogicas Brazil, the University of North
Carolina at Chapel Hill, Michigan State University, and the National
Optical Astronomy Observatory.
This research used data obtained with \xmm, an ESA science
mission with instruments and contributions directly funded by ESA
Member States and the USA (NASA).
This research used data from the Two Micron All Sky Survey, a
joint project of the University of Massachusetts and the Infrared
Processing and Analysis Center/California Institute of Technology,
funded by the National Aeronautics and Space Administration and the
National Science Foundation.
This research used data from the NASA/IPAC Extragalactic Database
(NED) which is operated by the Jet Propulsion Laboratory, California
Institute of Technology, under contract with the National Aeronautics
and Space Administration.
This research used data from the Sloan Digital Sky Survey, funded by
the Alfred P. Sloan Foundation, the participating institutions, the
National Science Foundation, the U.S. Department of Energy, the
National Aeronautics and Space Administration, the Japanese
Monbukagakusho, the Max Planck Society, and the Higher Education
Funding Council for England.

\normalsize


\bibliography{apj-jour,radio}
\bibliographystyle{apj}

\clearpage

\begin{center}
\begin{deluxetable}{l l c r c   r c c c r }
\tabletypesize{\scriptsize}
\tablecaption{X-ray Properties
\label{tab.xray} 
}
\tablehead{
\colhead{Cluster} &\colhead{Alt name} &\colhead{z}   &\colhead{\rfive} &\colhead{$n_eh(z)^{-2}$} &\colhead{log($t_{cool}$)} &\colhead{log(\w)}       &\colhead{log($M_{cl}$)}  &\colhead{X peak coordinates} &\colhead{Offset} \\ 
                  &                   &              &\colhead{(kpc)}  &\colhead{(cm$^{-3}$)}    &\colhead{(log(yr))}       &\colhead{(log(\rfive))} &\colhead{(log($\msun$))} &\colhead{J2000}              &\colhead{(kpc)}  \\ 
\colhead{(1)}     &\colhead{(2)}      &\colhead{(3)} &\colhead{(4)}    &\colhead{(5)}            &\colhead{(6)}             &\colhead{(7)}           &\colhead{(8)}            &\colhead{(9)}                &\colhead{(10)}\\
}
\startdata
RXCJ0003.8$+$0203 & A2700     & 0.0924 & $ 876^{+ 5}_{- 5}$ & 0.0162$\pm$0.0007 &  9.47$\pm$0.02 & -2.49$\pm$0.12 & $14.321^{+0.008}_{-0.008}$ & 00:03:49.7 $+$02:03:58 &   5$\pm$ 6 \\ 
RXCJ0006.0$-$3443 & A2721     & 0.1147 & $1059^{+10}_{-10}$ & 0.0079$\pm$0.0008 &  9.82$\pm$0.02 & -1.89$\pm$0.05 & $14.577^{+0.012}_{-0.013}$ & 00:05:59.9 $-$34:43:23 &  14$\pm$ 8 \\ 
RXCJ0020.7$-$2542 & A0022     & 0.1410 & $1045^{+ 5}_{- 5}$ & 0.0085$\pm$0.0007 &  9.67$\pm$0.02 & -2.20$\pm$0.05 & $14.571^{+0.007}_{-0.007}$ & 00:20:42.2 $-$25:42:25 &  32$\pm$ 9 \\ 
RXCJ0049.4$-$2931 & S0084     & 0.1084 & $ 807^{+ 7}_{- 7}$ & 0.0182$\pm$0.0014 &  9.37$\pm$0.02 & -2.64$\pm$0.13 & $14.221^{+0.012}_{-0.012}$ & 00:49:23.0 $-$29:31:14 &   5$\pm$ 7 \\ 
RXCJ0145.0$-$5300 & A2941     & 0.1168 & $1089^{+ 6}_{- 6}$ & 0.0038$\pm$0.0005 &  9.92$\pm$0.02 & -1.52$\pm$0.02 & $14.614^{+0.008}_{-0.008}$ & 01:44:59.7 $-$53:01:03 &  23$\pm$ 8 \\ 
RXCJ0211.4$-$4017 & A2948     & 0.1008 & $ 685^{+ 3}_{- 3}$ & 0.0214$\pm$0.0008 &  9.23$\pm$0.02 & -2.34$\pm$0.09 & $14.003^{+0.007}_{-0.007}$ & 02:11:24.8 $-$40:17:28 &   1$\pm$ 7 \\ 
RXCJ0225.1$-$2928 & \nodata   & 0.0604 & $ 693^{+ 7}_{- 8}$ & 0.0038$\pm$0.0005 &  9.60$\pm$0.02 & -1.92$\pm$0.05 & $14.003^{+0.014}_{-0.015}$ & 02:25:09.3 $-$29:28:36 &   4$\pm$ 4 \\ 
RXCJ0345.7$-$4112 & S0384     & 0.0603 & $ 688^{+ 5}_{- 4}$ & 0.0547$\pm$0.0010 &  9.07$\pm$0.02 & -2.28$\pm$0.07 & $13.992^{+0.011}_{-0.008}$ & 03:45:46.2 $-$41:12:14 &   4$\pm$ 4 \\ 
RXCJ0547.6$-$3152 & A3364     & 0.1483 & $1133^{+ 5}_{- 5}$ & 0.0088$\pm$0.0005 &  9.65$\pm$0.02 & -2.15$\pm$0.03 & $14.680^{+0.007}_{-0.007}$ & 05:47:38.4 $-$31:52:12 &  39$\pm$10 \\ 
RXCJ0605.8$-$3518 & A3378     & 0.1392 & $1045^{+ 5}_{- 5}$ & 0.0739$\pm$0.0013 &  8.95$\pm$0.02 & -2.23$\pm$0.03 & $14.571^{+0.007}_{-0.007}$ & 06:05:54.2 $-$35:18:09 &   7$\pm$ 9 \\ 
RXCJ0616.8$-$4748 & \nodata   & 0.1164 & $ 939^{+ 5}_{- 6}$ & 0.0119$\pm$0.0006 &  9.64$\pm$0.02 & -1.88$\pm$0.05 & $14.421^{+0.008}_{-0.008}$ & 06:16:51.7 $-$47:47:40 &  14$\pm$ 8 \\ 
RXCJ0645.4$-$5413 & A3404     & 0.1644 & $1279^{+ 7}_{- 7}$ & 0.0245$\pm$0.0014 &  9.37$\pm$0.02 & -2.41$\pm$0.04 & $14.846^{+0.008}_{-0.008}$ & 06:45:29.3 $-$54:13:40 &  10$\pm$11 \\ 
RXCJ0821.8$+$0112 & A0653     & 0.0822 & $ 755^{+ 6}_{- 6}$ & 0.0133$\pm$0.0008 &  9.60$\pm$0.02 & -2.35$\pm$0.12 & $14.123^{+0.010}_{-0.011}$ & 08:21:50.9 $+$01:11:52 &   6$\pm$ 6 \\ 
RXCJ0958.3$-$1103 & A0907     & 0.1669 & $1077^{+18}_{-16}$ & 0.0503$\pm$0.0023 &  9.08$\pm$0.02 & -2.47$\pm$0.08 & $14.622^{+0.022}_{-0.020}$ & 09:58:22.3 $-$11:03:54 &  17$\pm$11 \\ 
RXCJ1044.5$-$0704 & A1084     & 0.1342 & $ 931^{+ 2}_{- 2}$ & 0.1015$\pm$0.0018 &  8.87$\pm$0.02 & -2.14$\pm$0.02 & $14.419^{+0.004}_{-0.004}$ & 10:44:33.0 $-$07:04:09 &   4$\pm$ 9 \\ 
RXCJ1141.4$-$1216 & A1348     & 0.1195 & $ 885^{+ 3}_{- 2}$ & 0.0701$\pm$0.0010 &  8.94$\pm$0.02 & -2.27$\pm$0.05 & $14.345^{+0.004}_{-0.004}$ & 11:41:24.4 $-$12:16:37 &   5$\pm$ 8 \\ 
RXCJ1236.7$-$3354 & A0700     & 0.0796 & $ 753^{+ 6}_{- 0}$ & 0.0125$\pm$0.0007 &  9.41$\pm$0.02 & -2.28$\pm$0.05 & $14.118^{+0.011}_{-0.002}$ & 12:36:41.3 $-$33:55:37 &   8$\pm$ 6 \\ 
RXCJ1302.8$-$0230 & A1663     & 0.0847 & $ 842^{+ 4}_{- 4}$ & 0.0347$\pm$0.0006 &  9.20$\pm$0.02 & -1.82$\pm$0.02 & $14.265^{+0.007}_{-0.007}$ & 13:02:53.3 $-$02:31:00 &  17$\pm$ 6 \\ 
RXCJ1311.4$-$0120 & A1689     & 0.1832 & $1319^{+ 4}_{- 4}$ & 0.0465$\pm$0.0011 &  9.16$\pm$0.02 & -2.40$\pm$0.03 & $14.893^{+0.004}_{-0.004}$ & 13:11:29.5 $-$01:20:28 &   1$\pm$12 \\ 
RXCJ1516.3$+$0005 & A2050     & 0.1181 & $ 989^{+ 3}_{- 3}$ & 0.0109$\pm$0.0006 &  9.61$\pm$0.02 & -2.43$\pm$0.05 & $14.490^{+0.005}_{-0.005}$ & 15:16:18.1 $+$00:05:28 &  16$\pm$ 8 \\ 
RXCJ1516.5$-$0056 & A2051     & 0.1198 & $ 927^{+ 6}_{- 5}$ & 0.0104$\pm$0.0008 &  9.66$\pm$0.02 & -1.75$\pm$0.03 & $14.405^{+0.009}_{-0.008}$ & 15:16:44.2 $-$00:58:12 &   6$\pm$ 8 \\ 
RXCJ2014.8$-$2430 & \nodata   & 0.1538 & $1155^{+ 4}_{- 4}$ & 0.1291$\pm$0.0023 &  8.74$\pm$0.02 & -2.24$\pm$0.02 & $14.707^{+0.005}_{-0.005}$ & 20:14:51.7 $-$24:30:20 &   5$\pm$10 \\ 
RXCJ2023.0$-$2056 & S0868     & 0.0564 & $ 739^{+ 6}_{- 6}$ & 0.0092$\pm$0.0009 &  9.61$\pm$0.02 & -1.78$\pm$0.04 & $14.084^{+0.011}_{-0.011}$ & 20:22:58.8 $-$20:56:56 &   4$\pm$ 4 \\ 
RXCJ2129.8$-$5048 & A3771     & 0.0796 & $ 900^{+ 7}_{- 8}$ & 0.0052$\pm$0.0005 &  9.91$\pm$0.02 & -1.38$\pm$0.17 & $14.351^{+0.011}_{-0.012}$ & 21:29:40.9 $-$50:48:55 &  51$\pm$ 6 \\ 
RXCJ2149.1$-$3041 & A3814     & 0.1184 & $ 886^{+ 4}_{- 4}$ & 0.0549$\pm$0.0011 &  8.92$\pm$0.02 & -2.47$\pm$0.06 & $14.347^{+0.006}_{-0.006}$ & 21:49:07.6 $-$30:42:05 &   4$\pm$ 8 \\ 
RXCJ2157.4$-$0747 & A2399     & 0.0579 & $ 751^{+ 4}_{- 4}$ & 0.0034$\pm$0.0003 & 10.02$\pm$0.02 & -0.97$\pm$0.97 & $14.106^{+0.009}_{-0.009}$ & 21:57:29.5 $-$07:47:55 &  12$\pm$ 4 \\ 
RXCJ2217.7$-$3543 & A3854     & 0.1486 & $1022^{+ 4}_{- 4}$ & 0.0182$\pm$0.0009 &  9.45$\pm$0.02 & -2.74$\pm$0.49 & $14.546^{+0.006}_{-0.006}$ & 22:17:45.5 $-$35:43:30 &  10$\pm$10 \\ 
RXCJ2218.6$-$3853 & A3856     & 0.1411 & $1130^{+ 7}_{- 8}$ & 0.0126$\pm$0.0010 &  9.50$\pm$0.02 & -1.81$\pm$0.01 & $14.673^{+0.009}_{-0.009}$ & 22:18:40.3 $-$38:54:06 &  29$\pm$ 9 \\ 
RXCJ2234.5$-$3744 & A3888     & 0.1510 & $1283^{+ 4}_{- 5}$ & 0.0063$\pm$0.0005 &  9.80$\pm$0.02 & -2.12$\pm$0.03 & $14.843^{+0.005}_{-0.005}$ & 22:34:27.1 $-$37:44:02 & 112$\pm$10 \\ 
RXCJ2319.6$-$7313 & A3992     & 0.0984 & $ 788^{+ 5}_{- 5}$ & 0.0571$\pm$0.0018 &  8.80$\pm$0.02 & -1.66$\pm$0.02 & $14.186^{+0.009}_{-0.009}$ & 23:19:40.2 $-$73:13:38 &   5$\pm$ 7 \\ 
\enddata
\end{deluxetable}

\tablecomments{
(1) Name of cluster in \rexcess\ catalog. 
(2) Alternate name of cluster.
(3) Redshift of the cluster. 
(4) The radius of the cluster enclosing a mean overdensity of 500
    times the critical density, found by \cite{croston08a} using the
    $M_{500}-Y_X$ relation of \cite{arnaud07a}.
(5) Gas density at 0.008~\rfive, from \cite{croston08a}.
(6) Gas cooling time at 0.03~\rfive, from \cite{croston08a}; we
    classify cool cores as $t_{cool}<2\times10^{9}\unit{yr}$ or
    $\log(t_{cool})<9.3$.
(7) Standard deviation of centroid shifts, from B{\"o}hringer\etal (in prep); 
    we classify disturbed clusters as \w$> 0.01$~\rfive.
(8) Total cluster mass, found by \cite{pratt09a} using the
    $M_{500}-Y_X$ relation of \cite{arnaud07a}.
(9) Coordinates of the peak of the X-ray emission, from
    B{\"o}hringer\etal in prep, with an uncertainty of 4$''$.
(10) Offset of the selected BCG from the peak X-ray position.
 }
\end{center}

\clearpage

\begin{center}
\tablewidth{0pt}
\begin{deluxetable}{l l l }
\tablecaption{Observations
\label{tab.obs} 
}
\tablehead{
\colhead{Date} & \colhead{Targets} & \colhead{Telescope Instrument} 
}
\startdata
2007 September 15  & RXCJ 2217, 2218, 2234, 2319                   & SOAR SOI \\ 
2007 October 11    & RXCJ 0345, 0547, 0605, 0645                   & SOAR SOI \\ 
2008 March 8       & RXCJ 0821, 0958, 1516.3, 1516.5               & SOAR SOI \\ 
2008 July 6        & RXCJ 2014, 2023, 2048, 2129, 2157             & SOAR SOI \\ 
2008 July 7        & RXCJ 0006, 0020, 0049, 0145, 0211, 0225       & SOAR SOI \\ 
2008 October 4     & RXCJ 2149                                     & SOAR Goodman \\ 
2008 November 2    & RXCJ 0003                                     & SOAR Goodman \\ 
2009 January 15-16 & RXCJ 1044                                     & Lowell Hall 42$''$ \\ 
2009 April 17      & RXCJ 0616, 1141                               & SOAR Goodman \\ 
2009 April 27      & RXCJ 1236, 1302                               & SOAR Goodman \\ 
\nodata            & RXCJ 1311                                     & SDSS \\ 
\enddata
\end{deluxetable}

\end{center}


\begin{center}
\begin{deluxetable}{l  c r  c  c r  c c }
\tablewidth{0pt}
\tablecaption{Optical Properties - metric
\label{tab.opt-met} 
}
\tablehead{
\colhead{BCG coord} &\colhead{$R_{met}$} &\multicolumn{2}{c}{$R_{met,abs}$}       &\colhead{$L_{BCG}$}       &\colhead{PA}    &\colhead{e}    &\colhead{$\mu$}              \\
\colhead{J2000}     &\colhead{(mag)}     &\colhead{observed} &\colhead{restframe} &\colhead{($\log(\lsun)$)} &\colhead{(deg)} &               &\colhead{(mag arcsec$^{-1}$)}   \\ 
\colhead{(1)}       &\colhead{(2)}       &\colhead{(3)}      &\colhead{(4)}       & \colhead{(5)}            &\colhead{(6)}   & \colhead{(7)} & \colhead{(8)}   \\
}
\startdata
00:03:49.6 $+$02:04:00 &  15.06 & -23.07 & -23.16 & 11.06$\pm$0.02 & -81 & 0.45 & 21.24 \\
00:05:59.6 $-$34:43:17 &  15.59 & -23.04 & -23.16 & 11.06$\pm$0.02 & -77 & 0.51 & 21.18 \\
00:20:43.1 $-$25:42:28 &  16.23 & -22.89 & -23.04 & 11.01$\pm$0.02 &  24 & 0.47 & 21.22 \\
00:49:22.8 $-$29:31:12 &  15.53 & -22.98 & -23.09 & 11.03$\pm$0.02 & -25 & 0.04 & 21.92 \\
01:44:58.9 $-$53:01:12 &  16.22 & -22.46 & -22.58 & 10.83$\pm$0.02 &  74 & 0.21 & 22.15 \\
02:11:24.8 $-$40:17:28 &  15.64 & -22.70 & -22.80 & 10.91$\pm$0.02 & -62 & 0.23 & 22.09 \\
02:25:09.0 $-$29:28:38 &  14.61 & -22.55 & -22.61 & 10.84$\pm$0.02 & -76 & 0.31 & 22.16 \\
03:45:46.0 $-$41:12:16 &  13.87 & -23.28 & -23.34 & 11.13$\pm$0.02 &  52 & 0.09 & 21.77 \\
05:47:37.7 $-$31:52:24 &  16.36 & -22.88 & -23.05 & 11.01$\pm$0.02 &  19 & 0.09 & 22.05 \\
06:05:53.9 $-$35:18:08 &  16.08 & -23.01 & -23.16 & 11.06$\pm$0.02 & -73 & 0.11 & 21.93 \\
06:16:51.7 $-$47:47:45 &  15.22 & -23.45 & -23.57 & 11.22$\pm$0.02 &  83 & 0.53 & 20.73 \\
06:45:29.5 $-$54:13:37 &  16.26 & -23.23 & -23.41 & 11.16$\pm$0.02 &  65 & 0.31 & 21.19 \\
08:21:50.7 $+$01:11:49 &  15.21 & -22.66 & -22.74 & 10.89$\pm$0.02 & -18 & 0.34 & 22.30 \\
09:58:22.0 $-$11:03:51 &  16.56 & -22.96 & -23.14 & 11.05$\pm$0.02 & -27 & 0.26 & 21.60 \\
10:44:32.9 $-$07:04:07 &  16.39 & -22.62 & -22.76 & 10.90$\pm$0.02 &   2 & 0.29 & 22.07 \\
11:41:24.2 $-$12:16:37 &  15.49 & -23.24 & -23.37 & 11.14$\pm$0.02 &  -7 & 0.27 & 21.40 \\
12:36:41.3 $-$33:55:32 &  14.91 & -22.87 & -22.95 & 10.97$\pm$0.02 & -16 & 0.13 & 22.14 \\
13:02:52.6 $-$02:30:59 &  15.15 & -22.78 & -22.86 & 10.94$\pm$0.02 &  78 & 0.22 & 22.16 \\
13:11:29.5 $-$01:20:28 &  16.63 & -23.11 & -23.31 & 11.12$\pm$0.02 &  58 & 0.13 & 21.49 \\
15:16:17.9 $+$00:05:21 &  15.87 & -22.83 & -22.95 & 10.98$\pm$0.02 &  43 & 0.41 & 21.57 \\
15:16:44.2 $-$00:58:09 &  15.80 & -22.94 & -23.06 & 11.02$\pm$0.02 & -45 & 0.29 & 21.59 \\
20:14:51.7 $-$24:30:22 &  16.10 & -23.33 & -23.51 & 11.20$\pm$0.02 &  37 & 0.21 & 21.38 \\
20:22:59.1 $-$20:56:56 &  14.14 & -22.86 & -22.91 & 10.96$\pm$0.02 & -66 & 0.09 & 21.95 \\
20:48:11.6 $-$17:49:03 &  15.83 & -23.40 & -23.56 & 11.22$\pm$0.02 &  16 & 0.14 & 21.45 \\
21:29:42.4 $-$50:49:26 &  15.10 & -22.69 & -22.76 & 10.90$\pm$0.02 &  10 & 0.19 & 22.20 \\
21:49:07.4 $-$30:42:05 &  15.29 & -23.42 & -23.55 & 11.21$\pm$0.02 &   0 & 0.19 & 21.36 \\
21:57:29.4 $-$07:47:44 &  14.45 & -22.61 & -22.67 & 10.86$\pm$0.02 &  80 & 0.42 & 22.55 \\
22:17:45.8 $-$35:43:29 &  16.00 & -23.25 & -23.41 & 11.16$\pm$0.02 & -22 & 0.20 & 21.11 \\
22:18:39.4 $-$38:54:02 &  16.03 & -23.09 & -23.24 & 11.09$\pm$0.02 & -35 & 0.36 & 21.24 \\
22:34:24.6 $-$37:43:31 &  16.47 & -22.81 & -22.98 & 10.98$\pm$0.02 &  18 & 0.15 & 21.84 \\
23:19:40.5 $-$73:13:36 &  16.39 & -21.88 & -21.99 & 10.59$\pm$0.02 &   0 & 0.26 & 22.72 \\
\enddata
\tablecomments{ All magnitudes and colors are on the AB scale
and corrected for Galactic dust extinction.  
(1) BCG coordinates
(2) R~band apparent magnitude in metric radius of
    12\unit{h^{-1}}~kpc, uncertainty of 0.05 mag.
(3,4) R~band absolute magnitude in metric radius of
       12\unit{h^{-1}}~kpc, uncertainty of 0.05 mag.  Column 4 is 
       converted to restframe. 
(5) Luminosity of BCG corresponding to $R_{met,abs}$ converted
    to restframe. 
(6,7,8) Position angle (degrees E of N), ellipticity, and corrected surface
           brightness of the isophote with semi-major axis
           12\unit{h^{-1}}~kpc.  
}
\end{deluxetable}
\end{center}

\clearpage

\begin{center}
\begin{deluxetable}{l  c r  c  c c  c    c  c c  c  r c c }
\tablewidth{0pt}
\tablecaption{Optical Properties - Isophotal
\label{tab.opt-iso} 
}
\tablehead{
\colhead{2MASS Name} & \colhead{Class} &\colhead{Aper}   &\colhead{$R_{K20}$} &\multicolumn{2}{c}{$R-K$ color}         &\colhead{$J-K$ color} &\colhead{SMA}   \\
                     &                 &\colhead{($''$)} &\colhead{(mag)}     &\colhead{observed} &\colhead{restframe} &\colhead{restframe}   &\colhead{(kpc)} \\
\colhead{(1)}        &\colhead{(2)}    &\colhead{(3)}    &\colhead{(4)}       & \colhead{(5)}     &\colhead{(6)}       &\colhead{(7)}         &\colhead{(8)}  \\
}
\startdata
2MASX J00034964$+$0203594 & 1 & 20.2 &  14.57 & 1.31 & 1.08$\pm$0.10 &  0.18$\pm$0.10 & 40 \\
2MASX J00055975$-$3443171 & 1 & 12.4 &  15.46 & 1.18 & 0.90$\pm$0.10 & -0.05$\pm$0.10 & 41 \\
2MASX J00204314$-$2542284 & 2 & 18.7 &  15.32 & 1.39 & 1.01$\pm$0.09 &  0.14$\pm$0.09 & 47 \\
2MASS J00492286$-$2931124 & 2 &\nodata &\nodata &\nodata &\nodata    & \nodata        & 30 \\
2MASX J01445891$-$5301110 & 3 & 10.6 &  16.01 & 1.17 & 0.88$\pm$0.13 &  0.15$\pm$0.16 & 25 \\
2MASX J02112484$-$4017261 & 1 & 15.3 &  15.33 & 1.23 & 0.98$\pm$0.11 & -0.03$\pm$0.12 & 28 \\
2MASS J02250904$-$2928383 & 3 &\nodata &\nodata &\nodata &\nodata    & \nodata        & 29 \\
2MASX J03454640$-$4112149 & 1 & 22.3 &  13.61 & 1.18 & 1.03$\pm$0.07 &  0.08$\pm$0.06 & 28 \\
2MASX J05473773$-$3152237 & 1 & 11.5 &  16.00 & 1.34 & 0.94$\pm$0.16 &  0.11$\pm$0.18 & 26 \\
2MASX J06055401$-$3518081 & 1 &  9.8 &  15.87 & 1.44 & 1.07$\pm$0.09 &  0.21$\pm$0.11 & 25 \\
2MASX J06165166$-$4747434 & 1 & 18.0 &  14.74 & 1.16 & 0.87$\pm$0.10 &  0.04$\pm$0.11 & 65 \\
2MASX J06452948$-$5413365 & 1 & 14.7 &  15.73 & 1.33 & 0.87$\pm$0.13 &  0.13$\pm$0.15 & 39 \\
2MASX J08215065$+$0111495 & 1 & 12.9 &  15.06 & 1.22 & 1.02$\pm$0.11 &  0.10$\pm$0.12 & 21 \\
2MASX J09582201$-$1103500 & 1 & 11.3 &  16.03 & 1.56 & 1.10$\pm$0.12 &  0.13$\pm$0.13 & 40 \\
2MASX J10443287$-$0704074 & 1 &  7.1 &  16.49 & 1.26 & 0.90$\pm$0.12 &  0.04$\pm$0.13 & 25 \\
2MASX J11412420$-$1216386 & 1 & 11.2 &  15.37 & 1.34 & 1.03$\pm$0.09 &  0.16$\pm$0.09 & 38 \\
2MASX J12364125$-$3355321 & 1 & 16.1 &  14.62 & 0.90 & 0.71$\pm$0.09 & -0.06$\pm$0.09 & 23 \\
2MASX J13025254$-$0230590 & 1 & 14.4 &  14.97 & 1.61 & 1.40$\pm$0.09 &  0.13$\pm$0.08 & 24 \\
2MASX J13112952$-$0120280 & 1 &  8.2 &  16.32 & 1.48 & 0.96$\pm$0.12 &  0.13$\pm$0.13 & 26 \\
2MASX J15161794$+$0005203 & 1 & 10.1 &  15.76 & 1.36 & 1.05$\pm$0.11 &  0.14$\pm$0.12 & 34 \\
2MASX J15164416$-$0058096 & 1 & 11.6 &  15.60 & 1.17 & 0.86$\pm$0.11 &  0.02$\pm$0.12 & 33 \\
2MASX J20145171$-$2430229 & 1 &  9.4 &  15.84 & 1.28 & 0.86$\pm$0.10 &  0.06$\pm$0.11 & 36 \\
2MASX J20225911$-$2056561 & 2 & 19.9 &  13.77 & 0.79 & 0.65$\pm$0.09 & -0.09$\pm$0.08 & 25 \\
2MASX J20481162$-$1749034 & 3 &  9.3 &  15.63 & 1.27 & 0.87$\pm$0.10 &  0.01$\pm$0.11 & 36 \\
2MASX J21294244$-$5049260 & 3 & 11.6 &  15.27 & 1.03 & 0.84$\pm$0.10 & -0.08$\pm$0.11 & 25 \\
2MASX J21490737$-$3042043 & 1 & 14.0 &  14.91 & 1.12 & 0.81$\pm$0.09 &  0.04$\pm$0.08 & 38 \\
2MASX J21572939$-$0747443 & 3 & 10.3 &  14.72 & 1.13 & 0.99$\pm$0.07 &  0.01$\pm$0.06 & 18 \\
2MASX J22174585$-$3543293 & 1 & 13.6 &  15.25 & 1.32 & 0.91$\pm$0.10 &  0.09$\pm$0.11 & 45 \\
2MASX J22183938$-$3854018 & 2 & 14.1 &  15.50 & 1.23 & 0.85$\pm$0.09 &  0.08$\pm$0.09 & 51 \\
2MASX J22342463$-$3743304 & 3 & 11.0 &  15.88 & 1.24 & 0.83$\pm$0.10 &  0.00$\pm$0.11 & 33 \\
2MASX J23194046$-$7313366 & 1 &  9.7 &  16.44 & 1.54 & 1.30$\pm$0.13 &  0.43$\pm$0.17 & 16 \\
\enddata
\tablecomments{ All magnitudes and colors (including 2MASS) are on the AB scale
and corrected for Galactic dust extinction.   
(1) 2MASS name of selected BCG (see \S\ref{special} for RXCJ0049 and RXCJ0225).  
(2) BCG classification (see \S\ref{opt.id}. 
(3) Semi-major axis of aperture used for $R-K$ color, \ie\ the 2MASSX isophote
    where $K=20$ mag\unit{arcsec^{-2}}.
(4) R~band apparent magnitude in the 2MASSX K20 aperture, with
    masking to match 2MASSX, uncertainty 0.05 mag.
(5,6) $R-K$ color, both~bands measured in same aperture. 
      Column 6 is converted to restframe. 
(7) 2MASS $J-K$ color, converted to restframe.
(8) Semi-major axis of the isophote at 22.5 mag\unit{arcsec^{-2}}.
}
\end{deluxetable}
\end{center}


\begin{center}
\begin{deluxetable}{l  l l  l l  l l }
\tablewidth{0pt}
\tablecaption{Correlation statistics
\label{tab.stats} 
}
\tablehead{
\colhead{Parameters} & \multicolumn{2}{c}{CC Clusters} & \multicolumn{2}{c}{Non-CC Clusters} & \multicolumn{2}{c}{All Clusters} \\
                    & \colhead{R} &\colhead{P}     & \colhead{R} &\colhead{P}  & \colhead{R} &\colhead{P}  \\
}
\startdata
{\bf Optical - Optical}    &             &         &             &        &             &         \\
$R-K$ vs. $L_{BCG}$   & \phnm 0.41  &  0.24   & \phnm 0.12  & 0.64   & \phnm 0.13  & 0.49    \\
$R-K$ vs. $J-K$       & \phnm 0.73  &  0.016  & \phnm 0.70  & 0.00078 & \phnm 0.70  & 0.000020 \\[6pt]
{\bf X-ray - X-ray}        &             &         &             &        &             &         \\
$n_e$ vs. $M_{cl}$    & \phnm 0.37  &  0.29   & \phnm 0.095 & 0.70   & \phnm 0.067 & 0.73    \\
$n_e$ vs. \w          & \phnm 0.33  &  0.35   &    $-$0.74  & 0.00026 &    $-$0.56  & 0.0017  \\[6pt]
{\bf X-ray - Optical }     &             &         &             &        &             &         \\
$M_{cl}$ vs. $L_{BCG}$& \phnm 0.16  &  0.65   & \phnm 0.46  & 0.047  & \phnm 0.38  & 0.040    \\
$n_e$  vs. $e$        &    $-$0.091 &  0.80   &    $-$0.067 & 0.78   &    $-$0.26  & 0.17    \\
\w    vs. $L_{BCG}$   &    $-$0.37  &  0.29   &    $-$0.44  & 0.060  &    $-$0.46  & 0.012   \\
$n_e$  vs. $L_{BCG}$  & \phnm 0.36  &  0.31   & \phnm 0.67  & 0.0016 & \phnm 0.56  & 0.0017  \\
\enddata
\tablecomments{ R is the Spearman rank-order correlation coefficient.
P is the probability that the parameters are not correlated.  Correlations
with the offset between the BCG and X-ray peak are not listed because
many offsets are consistent with zero.  Clusters RXCJ2319 and RXCJ2048 have
been left out of all statistics, see \S\ref{special}.   
}
\end{deluxetable}
\end{center}

\clearpage

\end{document}